\documentclass[12pt,a4paper]{article}
\usepackage{color,graphics,epsfig,amsmath,amsfonts,citesort}

\begin{document}


\setlength{\unitlength}{1mm}
\def\noi{\noindent}
\def\lsim{\;\raise0.3ex\hbox{$<$\kern-0.75em\raise-1.1ex\hbox{$\sim$}}\;}
\def\gsim{\;\raise0.3ex\hbox{$>$\kern-0.75em\raise-1.1ex\hbox{$\sim$}}\;}
\def\bce{\begin{center}}
\def\ece{\end{center}}
\def\bea{\begin{eqnarray}}
\def\eea{\end{eqnarray}}
\def\beq{\begin{equation}}
\def\eeq{\end{equation}}
\def\ba{\begin{array}}
\def\ea{\end{array}}
\def\nn{\nonumber}
\def\micro{{\tt micrOMEGAs}}
\def\calc{{\tt CalcHEP}}
\def\comp{{\tt CompHEP}}
\def\nmh{{\sc nmhdecay}}
\def\nmtools{{\tt NMSSMTools}}
\def\susy{{\sc susy}}
\def\wh{\widehat}
\def\wt{\widetilde}
\def\d{\delta}
\def\b{\beta}
\def\D{\Delta}
\def\e{\epsilon}
\def\g{\gamma}
\def\G{\Gamma}
\def\l{\lambda}
\def\k{\kappa}
\def\t{\theta}
\def\s{\sigma}
\def\S{\Sigma}
\def\x{\chi}
\def\sf{\wt f}
\def\bino{{\wt B}}
\def\wino{{\wt W}}
\def\higgsino{{\wt H}}
\def\singlino{{\wt S}}
\def\tb{\tan\!\b}
\def\ctb{\cot\!\b}
\def\cb{\cos\!\b}
\def\sb{\sin\!\b}
\def\sbb{\sin\!2\b}
\def\tw{\tan\!\t_W}
\def\cw{\cos\!\t_W}
\def\sw{\sin\!\t_W}
\def\Oh2{\Omega h^2}
\def\mhf{M_{1/2}}
\def\m{m_0}
\def\A{A_0}
\def\Ak{A_\kappa}
\def\Al{A_\lambda}
\def\ma{m_A}
\def\Msusy{M_{\rm susy}}
\def\ZZ{\hbox {\it Z\hskip -4.pt Z}}

\renewcommand{\theequation}{\thesection.\arabic{equation}}
\makeatletter
\@addtoreset{equation}{section}
\makeatother


\bce
{\Large\bf Dark Matter in a Constrained NMSSM} \\[6mm]
{\large C.~Hugonie$^1$, G.~B\'elanger$^2$,  A.~Pukhov$^3$} \\[4mm]
{\it\small
1) Laboratoire Physique Th\'eorique et Astroparticules \\
Unit\'e mixte de Recherche -- CNRS -- UMR5207 \\
Universit\'e de Montpellier II, F-34095 Montpellier, France\\
2) Laboratoire d'Annecy-le-Vieux de Physique Th\'eorique \\
Unit\'e mixte de Recherche -- CNRS -- UMR5108 \\
Universit\'e de Savoie, F-74941 Annecy-le-Vieux, France \\
3) Skobeltsyn Institute of Nuclear Physics \\
Moscow State University, 119992 Moscow, Russia
}\\[4mm]
\today
\ece

\begin{abstract}
We explore the parameter space of a Constrained Next-to-Minimal Supersymmetric
Standard Model with GUT scale boundary conditions (CNMSSM), and find regions where the
relic density of the lightest neutralino is compatible with the WMAP measurement. We
emphasize differences with the MSSM: cases where annihilation of the LSP occurs via a
Higgs resonance at low values of $\tb$ and cases where the LSP has a large singlino
component. The particle spectrum as well as theoretical and collider constraints are
calculated with \nmtools. All neutralino annihilation and coannihilation processes are
then computed with \micro, taking into account higher order corrections to the Higgs
sector.
\end{abstract}

\section{Introduction}

One of the attractive features of supersymmetric (\susy) extensions of the Standard
Model (SM) with conserved R-parity, is the presence of a good Dark Matter (DM)
candidate, the lightest \susy\ particle (LSP). Over the years, numerous studies have
examined the constraints on the parameter space of the constrained minimal
supersymmetric standard model (CMSSM) including that originating from the cold
DM abundance extracted from precision cosmological measurements, notably those
of WMAP~\cite{Spergel:2003cb} and SDSS~\cite{Tegmark:2003ud}.
All concluded that the CMSSM could in some region of
parameter space provide a satisfactory DM
candidate~\cite{Ellis:2003cw,Baer:2003yh,Chattopadhyay:2003xi,Profumo:2004at,Baltz:2004aw,Allanach:2006cc,Roszkowski:2007fd,Ellis:2007fu}.
The MSSM, or its constrained version, however face a naturalness problem -- the
so-called $\mu$ problem. The NMSSM is a simple extension of the MSSM that solves
this problem elegantly via the introduction of a gauge singlet superfield, S. The
effective $\mu$ parameter term is determined by the vev of this singlet field, which is
naturally of EW
scale~\cite{Nilles:1982mp,Frere:1983ag,Derendinger:1983bz,Ellis:1988er,Drees:1988fc,Ellwanger:1993xa,Ellwanger:1996gw,King:1995vk,Franke:1995tc}.

The NMSSM contains an extra scalar and pseudoscalar states in the Higgs sector as well
as an additional neutralino, the singlino. Owing both to modifications in the Higgs
sector and the neutralino sector, the DM properties can differ from those of the
MSSM~\cite{Flores:1990bt,Stephan:1997rv,Stephan:1997ds,Belanger:2005kh,Gunion:2005rw,Cerdeno:2007sn}.
In particular a LSP with a large singlino component has different annihilation
properties than the bino LSP that is in general found in the CMSSM. It is even possible
to have a very light singlino LSP if accompanied by a very light
scalar. Even when the LSP has no singlino component, the more
elaborate Higgs sector of the model provides additional channels for rapid annihilation
through Higgs exchange. This can have implications for direct and indirect detection
rates~\cite{Flores:1991rx,Cerdeno:2004xw,Cerdeno:2007sn,Ferrer:2006hy}.

The purpose of this paper is to explore a  constrained version of the NMSSM
(CNMSSM) with semi-universal parameters defined at the GUT scale. Our choice
of semi-universality (i.e. non universal singlet sector) rather than strict universality
is motivated in sec.~\ref{sec:model}. In this first analysis we emphasize the
differences with the CMSSM predictions as concerns the neutralino properties and
its annihilation. We will therefore concentrate on regions of parameter space where
the LSP has some singlino component as well as regions where the spectrum is
such that annihilation can take place near a Higgs resonance. In the former case, we
will see that coannihilation processes, with staus or other neutralinos play a crucial
role. Of course, in large regions of parameter space we expect to recover features of
the CMSSM with a LSP that is mostly bino and can only annihilate efficiently near a
Higgs resonance or coannihilate with light sleptons. In this case the only difference
with the CMSSM will be additional allowed parameter space due to relaxed constraints
from LEP on the Higgs sector.

To evaluate the supersymmetric spectrum we use the NMSPEC program from the\linebreak
\nmtools\ package~\cite{Ellwanger:2006rn}. Using renormalization group equations (RGEs)
and starting from GUT scale parameters, this code computes the Higgs spectrum including
higher order corrections as well as the masses of sparticles at one-loop. It also checks
all the available collider constraints as well as various theoretical constraints. For
the computation of the relic density of DM we rely on \micro~\cite{Belanger:2006is}
which is included in \nmtools.

The paper is organized as follows: in sec.~\ref{sec:model} we briefly describe the model.
In sec.~\ref{sec:DM} we discuss the main channels for annihilation. In sec.~\ref{sec:res}
we present typical case studies. Conclusions follow in sec.~\ref{sec:dis}.

\section{The CNMSSM}\label{sec:model}

In the present paper we discuss the NMSSM with a scale invariant superpotential
\beq
W = \l {S} {H}_u {H}_d + \frac{\k}{3} \, {S}^3 + ({\rm Yukawa\ couplings}) \, ,
\eeq
where the weak scale originates from the soft \susy\ breaking scale only, i.e. where
no supersymmetric dimensionful parameters as $\mu$ are present in the superpotential.
The soft \susy\ breaking terms in the Higgs sector are then given by
\bea
V_{\rm soft} & = & m_{H_u}^2 |H_u|^2 + m_{H_d}^2 |H_d|^2 + m_S^2 |S|^2 \nn \\
& & + \left( \l \Al H_u H_d S + \frac{1}{3}\k \Ak S^3 + \rm{h.c.} \right) \, .
\eea

A possible cosmological domain wall problem~\cite{Abel:1995wk}
caused by a global $\ZZ_3$ symmetry can be avoided by introducing
suitable non-renormalizable operators~\cite{Abel:1996cr,Panagiotakopoulos:1998yw}
which neither generate dangerously large singlet tadpole
diagrams~\cite{Nilles:1982mp,Ellwanger:1983mg,Bagger:1995ay,Bagger:1993ji}
nor affect the low energy phenomeno\-logy.
Other models solving the MSSM $\mu$ problem with extra gauge singlet fields and with
no domain walls include the nearly minimal \susy\ model (nMSSM), with an additional
tadpole term for the singlet in the superpotential and/or in the soft scalar
potential~\cite{Panagiotakopoulos:1999ah,Panagiotakopoulos:2000wp,Dedes:2000jp},
and the UMSSM, with an extra $U(1)'$ gauge
symmetry~\cite{Suematsu:1994qm,Cvetic:1997ky,Demir:2005ti,Barger:2006dh}.
The DM properties as well as the possibility of generating the baryon
asymmetry of the universe within these models have been analysed in
refs.~\cite{Hugonie:2003yu,Menon:2004wv,Huber:2006wf,Balazs:2007pf,de Carlos:1997yv,Suematsu:2005bc,Nakamura:2006ht,Kang:2004pp,Barger:2005hb,Barger:2007nv}.

Constraining the parameters of the NMSSM by imposing universality at the
GUT scale is not as direct as in the MSSM. In the CMSSM, the free parameters
are the universal GUT scale soft terms $\A$, $\m$ and $\mhf$.
In addition, $M_Z$ and $\tb$ (at the weak scale)
are used as inputs, and the two minimization equations of the Higgs
potential w.r.t. the two real Higgs vevs $h_u$ and $h_d$ are used to
compute $\mu$ and $B$ in terms of the other parameters
(this leaves the sign of $\mu$ as a free parameter).
Both $\mu$ and $B$ have only a small effect on the
RGEs of the other parameters (via threshold effects from
particles whose masses depend on $\mu$ and/or $B$).
In numerical codes, this is usually solved by an iterative procedure.

At first sight, an application of this procedure to the NMSSM is not possible:
neither $\mu$ nor $B$ are present and one has to cope with three coupled
minimization equations w.r.t. the Higgs vevs $h_u$, $h_d$ and $s$. This
means that starting from strict universality, with $\mhf$, $\m$, $\A$ as well
as $\l$ and $\k$ as free parameters at the GUT scale, one usually ends up
with the wrong value of $M_Z$ at the weak scale (as no dimensionful parameter
is left to tune the correct value). In addition, $\tb$ cannot be a free
parameter in this approach. To solve this problem, the first phenomenological studies of
the NMSSM~\cite{Ellwanger:1993xa,Ellwanger:1996gw,King:1995vk,Ellwanger:1997jj}
used the fact that $\mhf$ can be factorised out of the RGEs and took  universal "scaled"
parameters at the GUT scale as inputs: $\A/\mhf$ and $\m/\mhf$, as well
as $\l$ and $\k$. In this approach however, $\tb$ and $\mhf$ are output parameters,
computed from the minimization of the potential and the known value of $M_Z$.
Furthermore, one usually finds that in the strictly universal NMSSM $\A, \m \ll
\mhf$~\cite{Ellwanger:1997jj}. It is therefore difficult to extend the CMSSM DM
studies, usually presented as plots in the $\m,\mhf$ plane for fixed $\tb$. In this paper
we follow the semi-universal approach first used in the program
NMSPEC~\cite{Ellwanger:2006rn}, that we shall briefly explain now.

First, let us assume that $\l$ as well as all the soft terms (except $m_S^2$) are known
at the weak scale (e.g. after integration of the RGEs down from the GUT scale). One can
define effective ($s$ dependent) parameters at the weak scale:
\beq\label{eq:def}
\mu = \l s \, , \qquad \nu = \k s \, , \qquad B = \Al + \nu \, .
\eeq
(In the following, we will often use $\nu$ rather than $\k$).
The minimization equations w.r.t. $h_u$ and $h_d$ can then be solved
for the effective $\mu$ and $B$, as in the MSSM, in terms of the other
parameters (incl. $M_Z$ and $\tb$). From $\mu$ and $B$ one then deduce
(for $\l$ and $A_\l$ given) both $s$ and $\k$. Finally, from the minimization
equation w.r.t. $s$, one can easily obtain the soft singlet mass $m_S^2$
in terms of all other parameters. At tree level, the minimization equations giving
$\mu$ (up to a sign), $B$ and $m_S^2$ (i.e. $\k$, $s$ and $m_S^2$) read:
\bea\label{eq:min}
\mu^2 & = & \frac{m_{H_d}^2 - m_{H_u}^2 \tan^2\!\b}{\tan^2\!\b - 1} - \frac{1}{2}M_Z^2 \, , \nn \\
B & = & \frac{\sbb}{2\mu} \left( m_{H_u}^2 + m_{H_d}^2 + 2\mu^2 + \l^2(h_u^2+h_d^2) \right) \, , \\
m_S^2 & = &  \l^2\frac{h_u h_d}{\mu}(\Al + 2\nu) - \nu(\Ak + 2\nu) - \l^2(h_u^2+h_d^2) \, . \nn
\eea

The radiative corrections to the scalar potential show a weak dependence
on $s$, $\k$ and $m_S^2$ which can be included in the minimization
equations. These become non-linear in the parameters to solve for and
have therefore to be solved iteratively. The derived parameters $\k$ and
$m_S^2$ affect the RGEs of the other parameters not only through
threshold effects around $\Msusy$, but also through the $\beta$
functions. However, the numerical impact is relatively small such
that an iterative procedure converges quite rapidly again.

As $m_S^2$ is computed from the minimization equations, it is difficult to
find parameters such that it assumes the same value as the Higgs
doublet (or other scalar) soft masses squared at the GUT scale. On the other
hand, the mechanism for the generation of soft \susy\ breaking terms could
easily treat the singlet differently from the other non-singlet matter
multiplets~\cite{Brax:1994ae}. Hence, we assume the following free
parameters for the CNMSSM:
\begin{itemize}
\item $\tb$, sign($\mu$) at the weak scale;
\item $\l$ at the \susy\ scale;
\item universal soft terms $\mhf$, $\m$ and $\A$ at the GUT scale.
Exceptions are:
\begin{itemize}
\item $m_S^2$ at the GUT scale which is an output, as described above;
\item $\Ak$ at the GUT scale which is considered as an independent parameter.
\end{itemize}
\end{itemize}

The reason for considering $\Ak$ at the GUT scale as a free parameter is
twofold: first, $\Ak = \A$ usually leads to negative mass squared in the Higgs
sector. Second, if an underlying mechanism for the generation of the soft \susy\
breaking terms treats the singlet differently from the other matter fields (as it is
already assumed for $m_S^2$), this will also affect the coupling $\Ak$ which
involves the singlet only. Hence, in our semi-universal approach, $\k$, $s$ and
$m_S^2$ are computed from the minimization equations. This implies important
differences with the general NMSSM where all the soft parameters as well as
$\l$, $\k$ and $\tb$ are taken as free parameters at the weak scale, especially
in the singlet neutralino (singlino) sector as we shall see.

Not all choices of parameters are allowed in the CNMSSM. Some lead to
negative mass squared for scalar fields (Higgs or sfermions), others to
Landau poles below the GUT scale for the dimensionless couplings. The
\nmtools\ package also includes all the available experimental constraints from
LEP and Tevatron on sparticle and Higgs searches (for details on the exclusion
channels see refs.~\cite{Ellwanger:2006rn,Ellwanger:2004xm}). It is the aim
of this paper to find out which regions in the parameter space of the CNMSSM
can fulfill all the theoretical and experimental tests and at the same time provide
the correct amount of DM. In sec.~4, we will present quantitative results,
but let us first have a qualitative approach here.

In order to do this, it is helpful to have a look at the symmetric $3\times 3$ CP even
Higgs mass matrix in the basis $(H_u,H_d,S)$ at tree level:
\beq
{\cal M}_S^2 =
\left[ \ba{ccc}
g^2 h_u^2 + \mu B \displaystyle\frac{h_d}{h_u} &
(2\l^2 - g^2) h_u h_d - \mu B &
2\l h_u \mu - \l h_d (\Al + 2\nu) \\ &
g^2 h_d^2 + \mu B \displaystyle\frac{h_u}{h_d} &
2\l h_d \mu - \l h_u (\Al + 2\nu) \\ & &
\l^2 \Al \displaystyle\frac{h_u h_d}{\mu}\, + \nu (\Ak + 4 \nu)
\ea\right] \, .
\eeq
To a good approximation, the $2\times 2$ doublet subsector is diagonalized
by the angle $\beta$ which gives a light eigenstate $h$ with mass
\beq\label{eq:mh}
m_h = \left(\cos^2\! 2\b + \displaystyle\frac{\l^2}{g^2} \sin^2\! 2\b \right) M_Z^2
\eeq
and a heavy eigenstate $H$ with a mass $m_H\sim m_A$ close to the MSSM-like
CP odd state (the larger $m_A$, the better this approximation). In the NMSSM,
one can define $m_A^2$ as the diagonal doublet term in the CP odd $2\times 2$
mass matrix after the Goldstone mode has been dropped. At tree level, it has the
same expression as in the MSSM:
\beq
m_A^2 = \frac{2\mu B}{\sbb} \, .
\eeq

It is already well known that large values of $\l$ are not allowed in the NMSSM
as they would imply a Landau pole below the GUT
scale~\cite{Nilles:1982mp,Frere:1983ag,Derendinger:1983bz,Ellis:1988er,Drees:1988fc,Ellwanger:1993xa,Ellwanger:1996gw,King:1995vk,Franke:1995tc}.
This leads to an upper bound on $\l\lsim .7$ which depends on $\tb$. Values of
$\l\gsim .1$ are not always allowed in the CNMSSM: on the one hand $\l$ increases
the mass of the light state $h$. This increase is relevant, however, only for small $\tb$.
On the other hand, $\l$ induces mixings between the doublet and the singlet states.
Since the singlet state is typically heavier than $\sim 100$~GeV, this mixing reduces
the mass of the lightest eigenstate $h$ which will then often violate bounds from LEP.
In order to maximize the mass of the lightest CP even mass eigenstate,
this singlet-doublet mixing has to vanish (as described in~\cite{Ellwanger:2006rm}),
which implies a relation between $\mu$, $\nu$, $\Al$, $\l$ and $\tb$. This relation is
generally not satisfied within the CNMSSM; then -- at least for large
values of $\tb$ -- this mixing effect disallows values of $\l\gsim .1$.

When $\l$ is small, the mixings between  the scalar and pseudoscalar singlet and
the doublet states are small and their masses read, respectively:
\beq\label{eq:singlets}
m_S^2 = \nu (\Ak+4\nu) \, , \quad m_P^2 = -3 \nu \Ak \, .
\eeq
The parameter $\Ak$ being only slightly renormalized from the GUT scale down to
the \susy\ scale, eq.~(\ref{eq:singlets}) shows that the masses of the singlet states
are directly proportional to the value of $\Ak$ at the GUT scale. The condition that
both squared masses are positive together with eq.~(\ref{eq:def}) implies
\beq\label{eq:rel}
-4(B - \Al)^2\ \lsim\ \Ak (B - \Al) \ \lsim\ 0 \, .
\eeq
The parameter $B$, obtained from the minimization equations~(\ref{eq:min}) depends
on $\m$, $\mhf$, $\l$ and $\tb$, while $\Al$ depends on $\A$, $\l$ and $\tb$. This means
that, for sign($\mu$) positive (which we will always assume in the following), either
$\Ak > 0$ and $\A > \wt A_0(\m,\mhf,\l,\tb)$ or $\Ak < 0$ and $\A < \wt A_0(\m,\mhf,\l,\tb)$.
Moreover, for $\Ak > 0$ and $\tb$ moderate, large values of $\m$ or $\mhf$ (implying $B$
large and positive) lead to a negative mass squared for the pseudoscalar singlet and are
therefore disallowed. If $\tb$ is large, however, $B$ remains small up to large values for
$\m$ and $\mhf$ which are no longer excluded.
\pagebreak

Finally, in the neutralino sector, we have a symmetric $5\times 5$ mass matrix,
given in the basis $(\bino,\wino,\higgsino_1,\higgsino_2,\singlino)$ by:
\beq
{\cal M}_{\wt\chi^0} =
\left[ \begin{array}{ccccc}
M_1 & 0 & \displaystyle\frac{g_1 h_u}{\sqrt{2}} & - \displaystyle\frac{g_1 h_d}{\sqrt{2}} & 0
\vspace*{.5ex} \\
& M_2 & - \displaystyle\frac{g_2 h_u}{\sqrt{2}} & \displaystyle\frac{g_2 h_d}{\sqrt{2}} & 0
\vspace*{.5ex} \\
& & 0 & -\mu & -\lambda h_d
\vspace*{.5ex} \\
& & & 0 & -\lambda h_u
\vspace*{.5ex} \\
& & & & 2 \nu
\end{array} \right]
\eeq
When $\l \ll 1$, the singlino decouples from other neutralinos and its mass is
$m_\singlino = 2\nu$. It is difficult to guess the input parameters that lead to a
singlino LSP as $\nu$ is a derived parameter. Nevertheless, using
eq.~(\ref{eq:def}) one can rewrite $m_\singlino = 2(B-\Al)$.
One can see from eq.~(\ref{eq:min}) that for large values of $\tb$ $B$ is small and
the mass of the singlino is simply $m_\singlino \sim -2 \Al$, i.e. it depends mainly
on $\A$ and is insensitive to $\m$ and $\mhf$. Hence, the singlino will be the LSP
for large values of $\mhf$ (where the bino is heavy). For moderate values of $\tb$,
the singlino mass depends also on $B$, which grows like $\m$ and $\mhf$ and
has the same sign as $\mu$ (assumed positive). Therefore, if $\Ak < 0$ and $\A
\lsim \wt A_0$, $\nu$ is positive and a singlino LSP is likely to appear at small
values of $\m$ and $\mhf$. On the contrary, if $\Ak>0$ and $\A \gsim \wt A_0$,
$\nu$ is negative and the singlino will be the LSP at large
$\m$ or $\mhf$, where $\nu \to 0$. (We remind that in this case, $\m$ and $\mhf$
are bounded from above in order to have a positive mass squared for the
pseudoscalar singlet). For $\l \gsim .1$ the singlino mixes with the higgsinos and
it is more difficult to make qualitative statements.

\section{Relic density of DM}\label{sec:DM}

The relic density calculation follows the usual procedure of evaluating the thermally
averaged cross section for annihilation and coannihilation of the LSP and solving for the
density evolution equation numerically. To do so we use \micro~\cite{Belanger:2006is}
adapted for the NMSSM~\cite{Belanger:2005kh}. For this a model has been specified in the
\calc~\cite{Pukhov:2004ca} format with the help of {\tt LANHEP}~\cite{Semenov:2002jw}.
Given the set of input parameters, as specified in the SLHA2
format~\cite{Skands:2003cj,Allanach:2006fy},
the code then finds the LSP before generating with \calc\ all the matrix elements of all
relevant processes of annihilation and, when necessary, coannihilation. An automatic
procedure for looking for s-channel poles is incorporated into the program such that a
more precise integration routine can be used in the event one is close to a pole. One
issue that has to be treated with special care is the one of radiative corrections to the
Higgs bosons masses and couplings. Those are calculated in \nmtools\ and can be very
important. Since knowing the precise value of the Higgs mass is often crucial to a relic
density calculation, we want to use the loop-corrected masses even though \calc\ computes
only tree-level matrix elements. To deal with this, as was described
in~\cite{Belanger:2005kh}, we write a general effective potential involving two doublets
and a singlet which includes 10 effective parameters. Taking the radiatively corrected
Higgs masses and mixings angles provided my \nmtools, we extract the value of these
effective parameters. It is then a simple matter to derive the corresponding trilinear and
quartic couplings of the scalar sector. Note however that the effective potential includes
only leading operators, and that in practice the number of parameters that we can extract
from observables (masses and mixings) in the Higgs sector is limited. While these
operators should include the dominant corrections to the masses and vertices, this
procedure can fail or show too large sensitivity to a given parameter. A signal that there
might be a problem with the procedure is that one of the dimensionless effective operators
is larger than 1. A warning is issued in this case~\footnote{It also means that there
might be a mismatch in the Higgs to Higgs partial widths between \nmtools\ and \micro\
since the set of radiative corrections to the Higgs masses that have been calculated is
much more complete than the ones to the partial widths. In general the difference is $<10
\%$. For the purpose of computing the relic density, the Higgs to Higgs decays can play a
role only near a resonance where the total width is the relevant parameter. A correction
to one partial width is therefore not so crucial.}.

A complete calculation of the neutralino DM in the NMSSM
defined at the weak scale has shown that often the same mechanisms
as in the MSSM for neutralino annihilation are at work. In the
constrained version of the models, those can be classified as
\begin{itemize}
\item Annihilation of a bino LSP through light sfermion exchange;
this occurs at small values of $\m, \mhf$.
\item Annihilation near a Higgs resonance (heavy or light Higgses);
this requires some higgsino component. The light Higgs resonance
is found for values of $\mhf$ near/below the LEP exclusion bound
while the heavy Higgs resonance is found at large values of $\tb$.
\item Annihilation of a mixed bino-higgsino LSP; this occurs at very
large $\m$ in the so-called focus point region.
\item Coannihilation of a bino with sfermions; this occurs at small
values of $\m$ when the stau is the NLSP or in a very small
region at low $\m$, $\mhf$ where the stop is the NLSP. The latter is
possible only for large negative values of $A_0$.
\end{itemize}

Nevertheless in the NMSSM one finds important differences with the
MSSM: the extra Higgses open up the possibility of more resonant
annihilation and the singlino component of the neutralino LSP
 can alter the prediction for annihilation cross-sections. The new
 mechanism that can provide a DM candidate compatible
 with the WMAP results are

\begin{itemize}
\item Annihilation near a pseudoscalar singlet resonance, occurring
at any values of $\tb$ provided $\l \gsim .1$.
\item Coannihilation of a singlino LSP with sfermions for small values
of $\m$ and $\l \ll 1$.
\item Coannihilation of a singlino LSP with higgsino NLSP;
this occurs at large $\m$, $\l \ll 1$ and is more likely for large values of $\tb$.
\item Coannihilation of a singlino LSP with bino NLSP, the bino rapidly
annihilating through a Higgs resonance, for large $\tb$ and $\l \ll 1$.
\end{itemize}

In the following section we will give specific examples where the
old and new mechanisms for LSP (co)annihilation are at work.

\section{Results}\label{sec:res}

We first performed a general scan over the parameter space of the
CNMSSM for fixed values of the SM parameters in order to find
regions satisfying all collider constraints and compatible with
WMAP, that is $\Omega h^2< .129$~\footnote{The most recent WMAP
and SDSS results give tighter constraints on the relic density of
DM~\cite{Spergel:2006hy,Tegmark:2006az}. However adopting  a more
conservative approach when fitting this data, that is allowing
extra degrees of freedom in the cosmological model used~\cite{Lahav:2006qy},
 leads to the range $.094 < \Omega h^2< .136$~\cite{Hamann:2006pf}, giving results
similar to the ones we discuss below.}. We quote only the upper
bound since we assume that the neutralino does not necessarily
account for all the DM. We used the central value of the top quark
mass measured at the Tevatron, $m_t =
171.4$~GeV~\cite{Brubaker:2006xn}, and choose $m_b(m_b) =
4.214$~GeV and $\alpha_S(M_Z) = .1172$. Note that the top quark
mass does affect the value of the light Higgs mass as well as the
calculation of the spectrum at large $\m$ while $m_b$ affects the
Higgs masses and couplings especially at large $\tb$. As mentioned
in sec.~\ref{sec:model}, we assumed sign($\mu$) $> 0$.

We found 2 disconnected regions in the parameter space: One at small
$\tb \sim 2$ and large $\l \sim .5$, where the lightest Higgs boson is heavy
enough to pass the LEP constraints due to the specific NMSSM tree level
contribution to its mass as shown in eq.~(\ref{eq:mh}). The other allowed
region is for small $\l \lsim .1$ and $\tb \gsim 4$. In this case, the singlet
sector is almost decoupled from the rest of the theory and the bound on
the lightest Higgs doublet implies a lower bound on $\tb$ as in the CMSSM.

We then picked 7 couple of values for $\l$ and $\tb$: $\l = .5$, $\tb = 2$ in
the first region and $\l = .1$ or $.01$, $\tb = 5, 10$ or $50$ in the second.
For each couple of values of $\l$, $\tb$ we scanned randomly on the
remaining free parameters, namely $\m$, $\mhf$, $\A$, $\Ak$, in order to
find regions in the parameter space allowed by all theoretical and
experimental constraints. As mentioned in sec.~\ref{sec:model}, we
found either $\Ak < 0$ and $\A < \wt A_0(\l,\tb)$ or $\Ak > 0$ and $\A >
\wt A_0(\l,\tb)$, the latter case appearing only if $\l = .01$ and/or
$\tb = 50$ (the other values of $\l$, $\tb$ always lead to light states in the
Higgs sector excluded by LEP when $\Ak > 0$).

Finally, we identified the values of $\A$ and $\Ak$ for which the main
neutralino annihilation channel is a Higgs singlet resonance, or the LSP
is mainly singlino. For the former case we found either $\l = .5$, $\tb = 2$
and $\A \sim \Ak$ large and negative, or $\l = .1$, large $\A < 0$ and small
$\Ak < 0$. For $\l = .1$, $\tb = 50$ we also found singlet resonances at
large $\A > 0$ and small $\Ak > 0$. The singlino LSP scenario on the other
hand appears only for $\l = .01$. It requires $\Ak < 0$ and $\A \lsim \wt A_0
(\tb)$ or $\Ak > 0$ and $\A \gsim \wt A_0 (\tb)$. We will now present plots
in the $\m,\mhf$ plane for selected values of $\l$, $\tb$, $\A$ and $\Ak$.

\subsection{Large $\lambda$: singlet resonances}

We first consider the cases where $\l = .5$ or $.1$. Here, the LSP is mainly bino
and is not expected to have a large singlino component. The constraints from
Higgs searches at LEP can be very important, especially at small values of $\tb$,
when considering the rather low value of the top quark mass now measured at
Fermilab, $m_t = 171.4$~GeV.

\begin{figure}[t]
\hspace*{-2mm}\epsfig{file=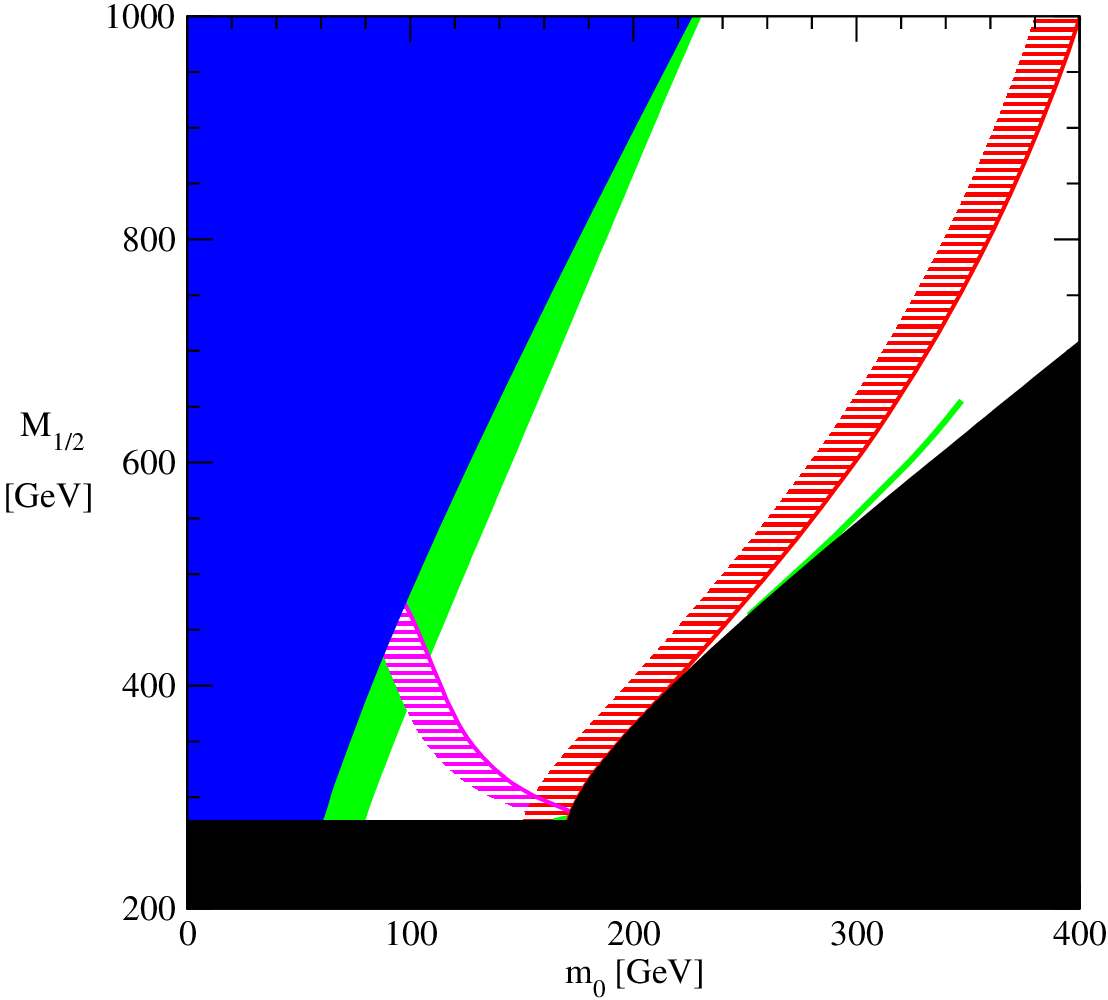, width=8cm}\hspace*{-6mm}(a)
\hspace*{6mm}\epsfig{file=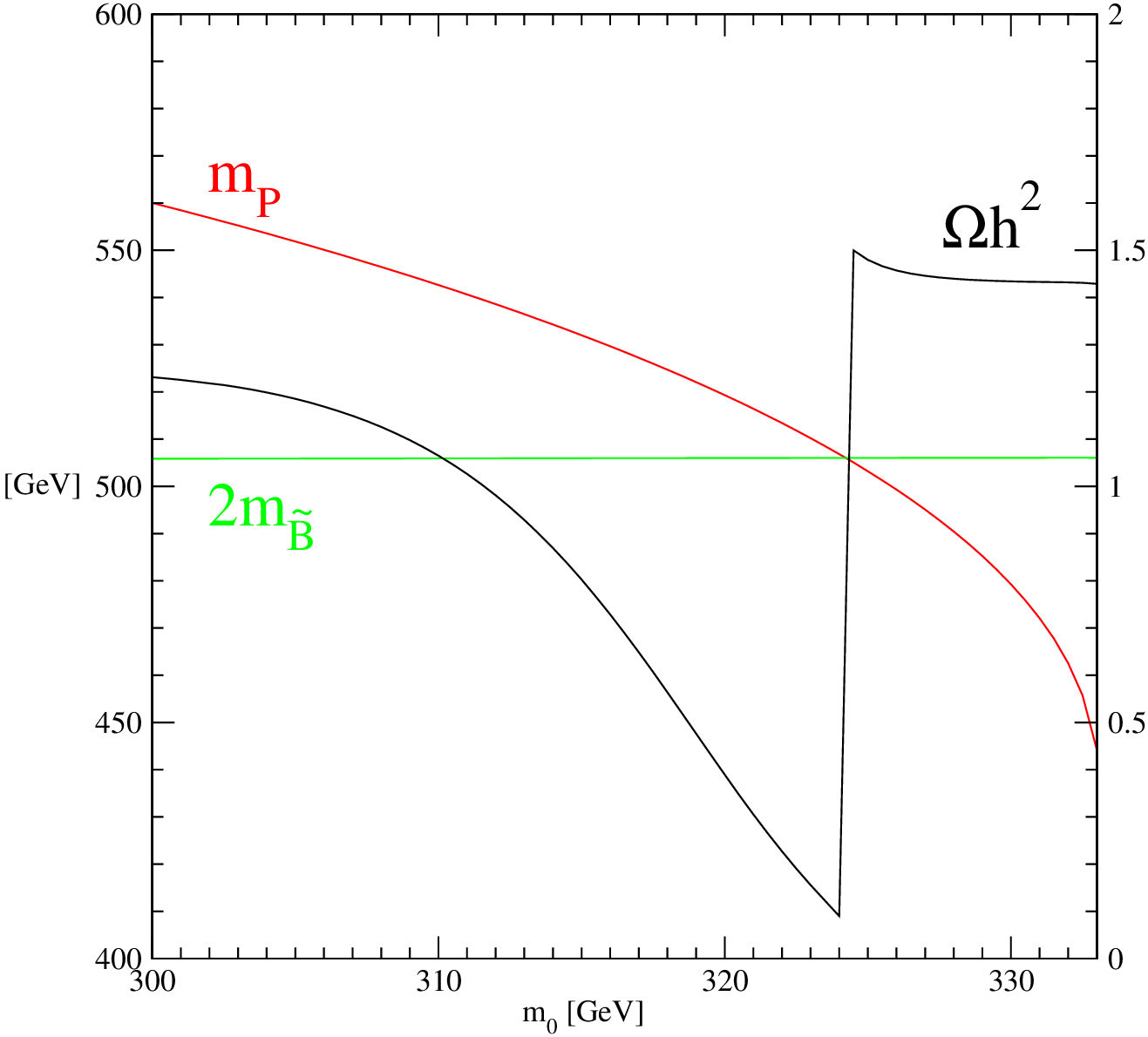, width=8cm}\hspace*{-6mm}(b)
\vspace*{-3mm}\caption{\small (a) The WMAP allowed region (green) in the $\m$, $\mhf$ plane for
$\l=.5$, $\tb=2$, $\A=-1300$~GeV, $\Ak=-1400$~GeV.
We show the region excluded by theoretical constraints or by LEP searches
on sparticles (black), the region where a sfermion -- the lightest stau here --
is the LSP (blue), the LEP limit from Higgs searches (red/hatch) and the contour
$m_h=111$~GeV (pink/hatch).
(b) The DM relic density $\Oh2$ (black), the pseudoscalar singlet mass $m_P$
(red) and twice the bino LSP mass $m_\bino$, as a function of $\m$ for the same
choice of parameters and $\mhf = 600$~GeV.}
\label{fig:tb2}
\end{figure}

\subsubsection{Small $\tb$}

The case $\tb=2$ is very characteristic of the NMSSM. Indeed, when
$\l$ is near its maximal value and $\tb$ is small, the light Higgs
doublet can be heavier than in the MSSM, as already mentioned, so
it is possible to satisfy the LEP constraints. Nevertheless, for
$\tb=2$ and $\l=.5$, large regions of parameter space are excluded
by experimental or theoretical constraints. For large negative
values of $\Ak$ and $\A$, however, one finds allowed regions in
the parameter space with the right order of magnitude for the
relic density of DM. For such constrained parameters, the singlino
always appear to be heavy while the LSP is mainly bino.
In some cases, the main bino
annihilation channel is a pseudoscalar singlet resonance. We show
in fig.~\ref{fig:tb2}(a) the various constraints in the $\m$,
$\mhf$ plane for $\A = -1300$~GeV and $\Ak = -1400$~GeV. The large
$\m$ region is theoretically excluded as well as the region $\mhf
\lsim 280$~GeV. Along the region where the lightest stau is the
LSP, at small $\m$, one finds a broad band where the DM relic
density is below the WMAP upper bound. In this band, the bino LSP
coannihilates with the stau (as in the CMSSM). Applying strictly
the LEP limits on Higgs searches from each decay channel, as it is
done in \nmtools, excludes a large fraction of the parameter
space, including this WMAP compatible band. Note however that
allowing for some theoretical uncertainty in the Higgs mass
calculation (estimated to be $\sim 3$~GeV) would restore most of
the forbidden parameter space. In particular, most of the WMAP
compatible  bino-stau coannihilation region has $m_h > 111$~GeV.
In the narrow region allowed by both theoretical and collider
constraints, one finds a thin band allowed by WMAP where the bino
LSP rapidly annihilates through a pseudoscalar singlet resonance.
In fig.~\ref{fig:tb2}(b) we show the DM relic density, the
pseudoscalar singlet mass and twice the bino LSP mass as a
function of $\m$ for the same choice of parameters as
fig.~\ref{fig:tb2}(a) and $\mhf = 600$~GeV. The sharp increase in the relic density
when $2m_{\tilde B}\approx m_P$ is typical of a pseudoscalar
resonance with a small width~\cite{Belanger:2005kh}.
This is to be contrasted with the
case of a narrow scalar resonance which shows smoother variations  because the scalar does not
contribute to the annihilation of a pair of neutralinos with  zero relative velocity. 

\subsubsection{Intermediate $\tb$}

\begin{figure}[t]
\hspace*{-2mm}\epsfig{file=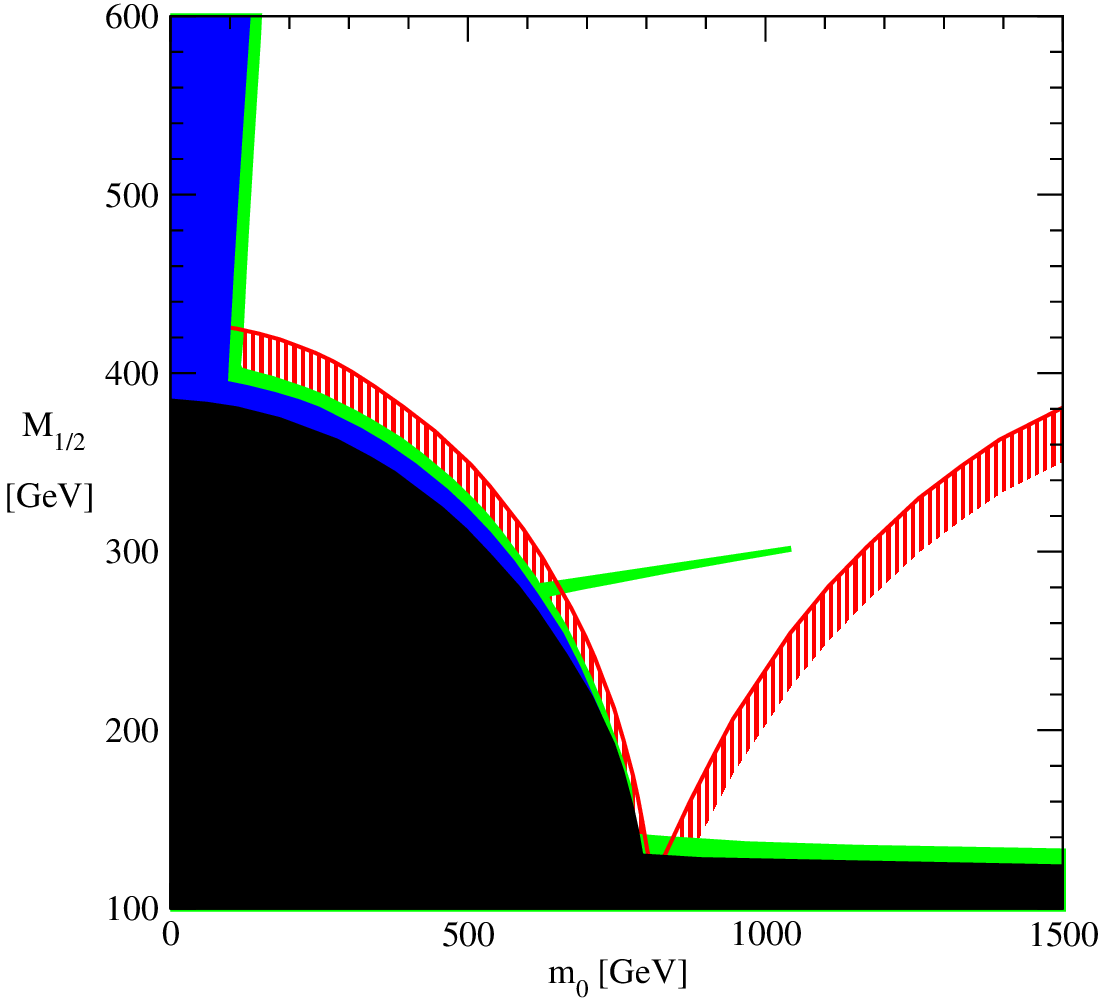, width=8cm}\hspace*{-6mm}(a)
\hspace*{6mm}\epsfig{file=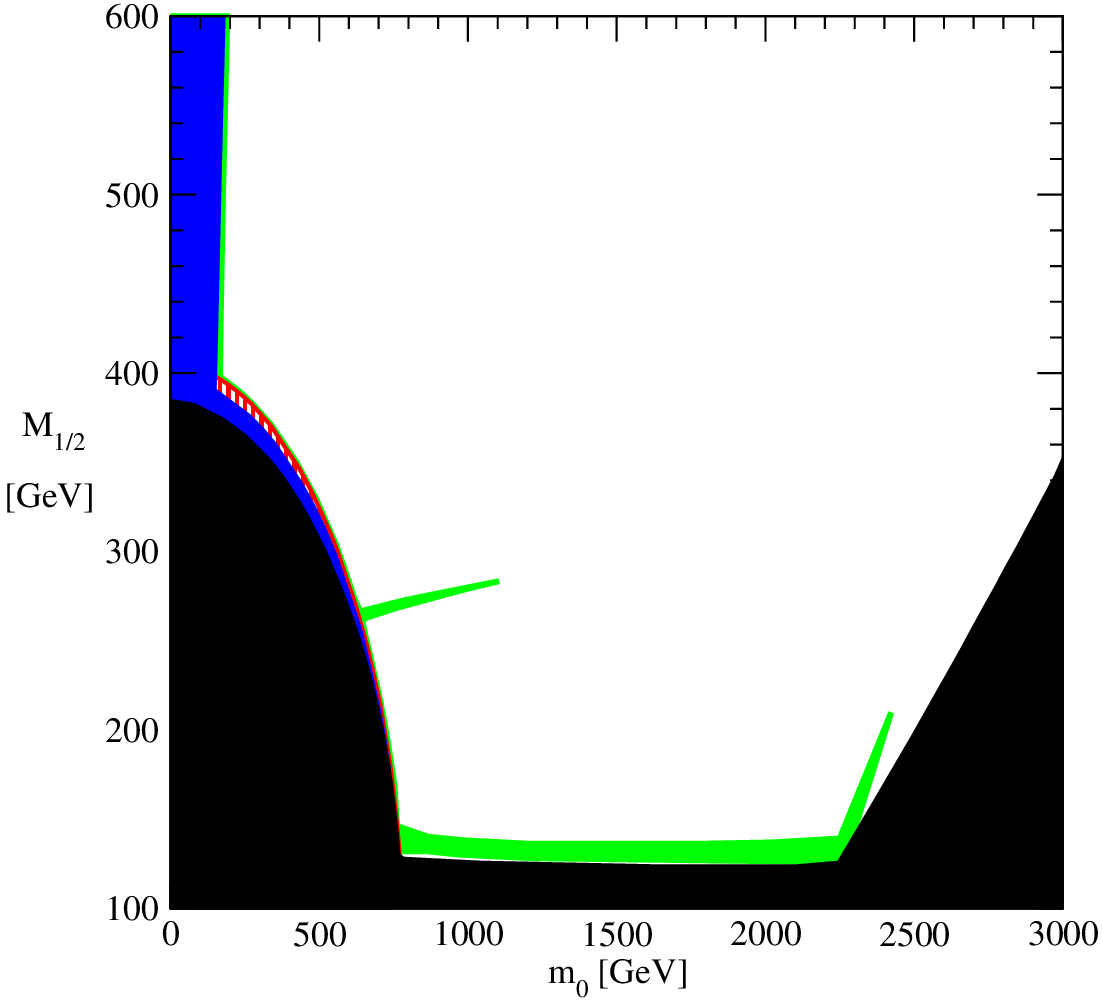, width=8cm}\hspace*{-6mm}(b)
\vspace*{-3mm}\caption{\small The WMAP allowed region in the $\m$, $\mhf$ plane for
(a) $\l=.1$, $\tb=5$, $\A=-1500$~GeV, $\Ak=-50$~GeV and
(b) $\l=.1$, $\tb=10$, $\A=-1500$~GeV, $\Ak=-50$~GeV.
Same color code as in fig.~\ref{fig:tb2}}
\label{fig:tb5_tb10}
\end{figure}

As explained in sec.~\ref{sec:model}, increasing $\tb$ implies $\l\lsim .1$,
larger values leading to a light scalar Higgs doublet excluded by LEP.
We now consider $\l=.1$, $\tb = 5$ or $10$. For these intermediate values of
$\tb$, the singlino can be light if $|\A|$ is close to its lower bound and $\m,\mhf$
are small. However, for our choice of $\l$ and $\tb$, small values of $|\A|$ always
lead to light states in the Higgs sector, excluded by LEP. Points for which $\Ak > 0$
are also excluded by light Higgs states. For $\Ak < 0$, and large $\A < 0$ one finds
points in agreement with LEP constraints for which the singlino is always heavy and
the LSP is mainly bino. In addition, small $\Ak < 0$ are favorable for bino LSP
annihilation through a Higgs resonance. The possible resonances are the
pseudoscalar singlet $P$ or the lightest scalar doublet $h$. As an example
we consider $\A=-1500$~GeV, $\Ak=-50$~GeV, see fig.~\ref{fig:tb5_tb10}.

The possibility of annihilation through a pseudoscalar resonance at
intermediate values of $\tb$ is a characteristic feature of the NMSSM.
In our case study, the pseudoscalar singlet resonance is found around
$\mhf\sim 300$~GeV, and corresponds to $|2m_\bino-m_P| \lsim 3$~GeV.
For larger negative values of $\Ak$, $m_P$ increases, as can be seen from
eq.~(\ref{eq:singlets}), which means that rapid annihilation would be possible
for a heavier bino LSP, that is a larger $\mhf$. However, when $m_\bino$ is large
it becomes increasingly difficult to rely exclusively on Higgs exchange to have
efficient enough annihilation. Thus, for large negative values of  $\Ak$, the
rapid annihilation region disappears. The pseudoscalar singlet exchange is
also dominant in the small region compatible with WMAP around $\m \sim
2.3$~TeV, see fig.~\ref{fig:tb5_tb10}(b). When annihilation is dominated by
the pseudoscalar singlet exchange, the annihilation channels are purely into
$bb$ and $\tau\tau$ pairs for light pseudoscalars, whereas the $tt$ channel
can contribute significantly, once passed the top threshold.

Rapid annihilation of the bino LSP through the light scalar doublet $h$ can
also occur at low values of $\mhf \sim 130$~ÊGeV. Note however that, as in
the CMSSM, the small $\mhf$ region is constrained by chargino searches
and Higgs searches at LEP, especially at small $\tb$. For $\tb=5$, the Higgs
constraint rules out practically all the scalar doublet annihilation region.
For $\tb=10$, the Higgs bound is relaxed, and both the WMAP and the
Higgs constraints are satisfied for $\mhf \sim 130$~GeV.
Alternatively, one could have increased $|\A|$ to relax the Higgs bound.

Sfermion coannihilation can also provide a mechanism to lower the
relic density below the WMAP upper bound. As in the MSSM, we find
a stau coannihilation band for values of $\m$ just above the stau LSP
forbidden region as well as a narrow stop coannihilation region at small
values of $\m$ and $\mhf$ just above the stop LSP forbidden region.
Note that the latter can only be found for large negative values of $\A$
and it satisfies the LEP constraints on the Higgs sector only when $\tb=10$.
Smaller values of $\m,\mhf$ are excluded as they lead to a negative
squared mass in the stop sector.

\subsubsection{Large $\tb$}

We next consider the case $\tb=50$, $\l=.1$. First note that one can
find allowed regions for $\A, \Ak < 0$, as we had for intermediate
values of $\tb$, but also for $\A, \Ak > 0$.
As explained in sec.~\ref{sec:model}, for large values of $\tb$, the
singlino mass is $-2 \Al$ which depends mainly on $\A$ but not on
$\m$ or $\mhf$. Hence, the singlino can be the LSP for large values
of $\mhf$, where the bino is heavy. However, since $\l=.1$, small
values of $|\A|$ lead to light Higgs states excluded by LEP. Hence, 
$|\A|$ has to be large and the regions where the singlino is the LSP 
are located at $\mhf \sim$ several TeV, where no mechanism is
available for singlino annihilation. 

\begin{figure}[t]
\hspace*{-2mm}\epsfig{file=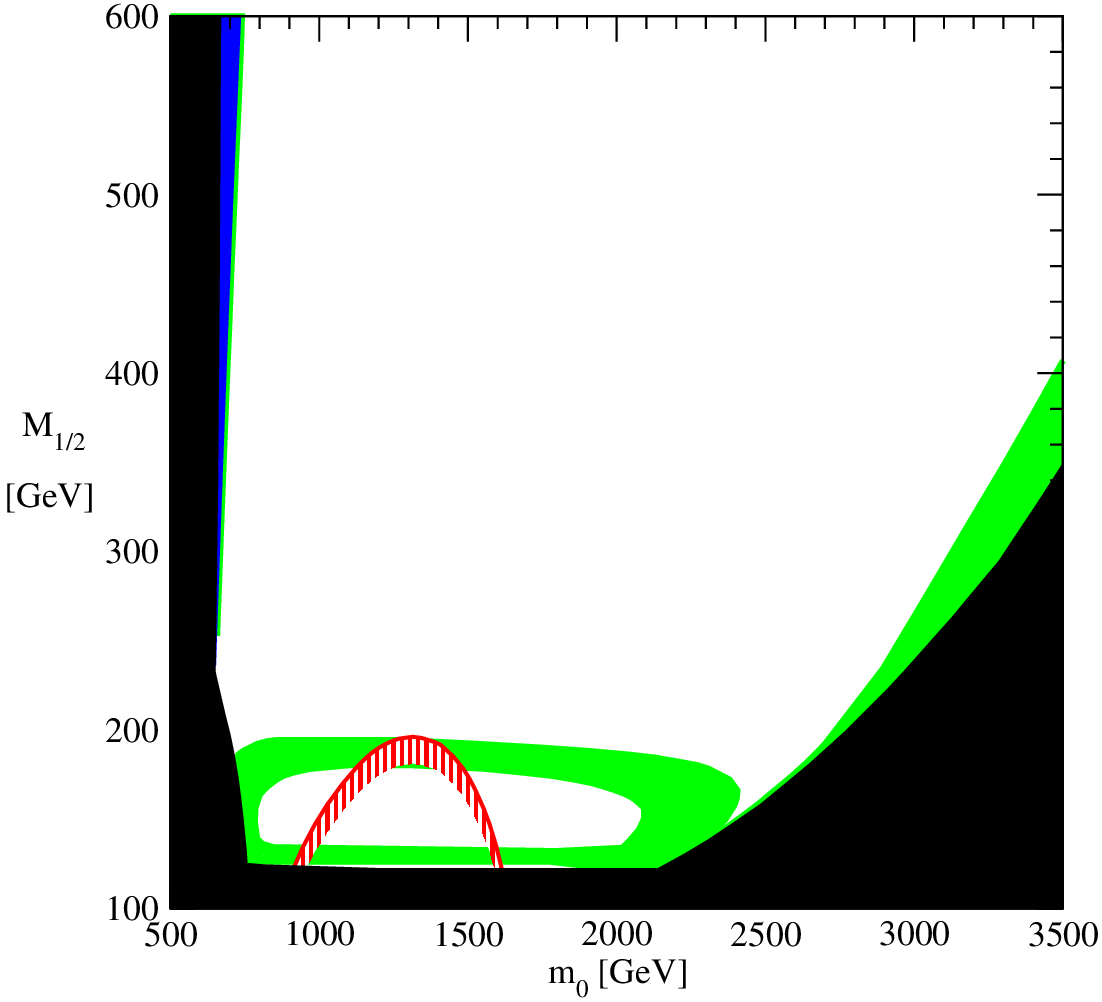, width=8cm}\hspace*{-6mm}(a)
\hspace*{6mm}\epsfig{file=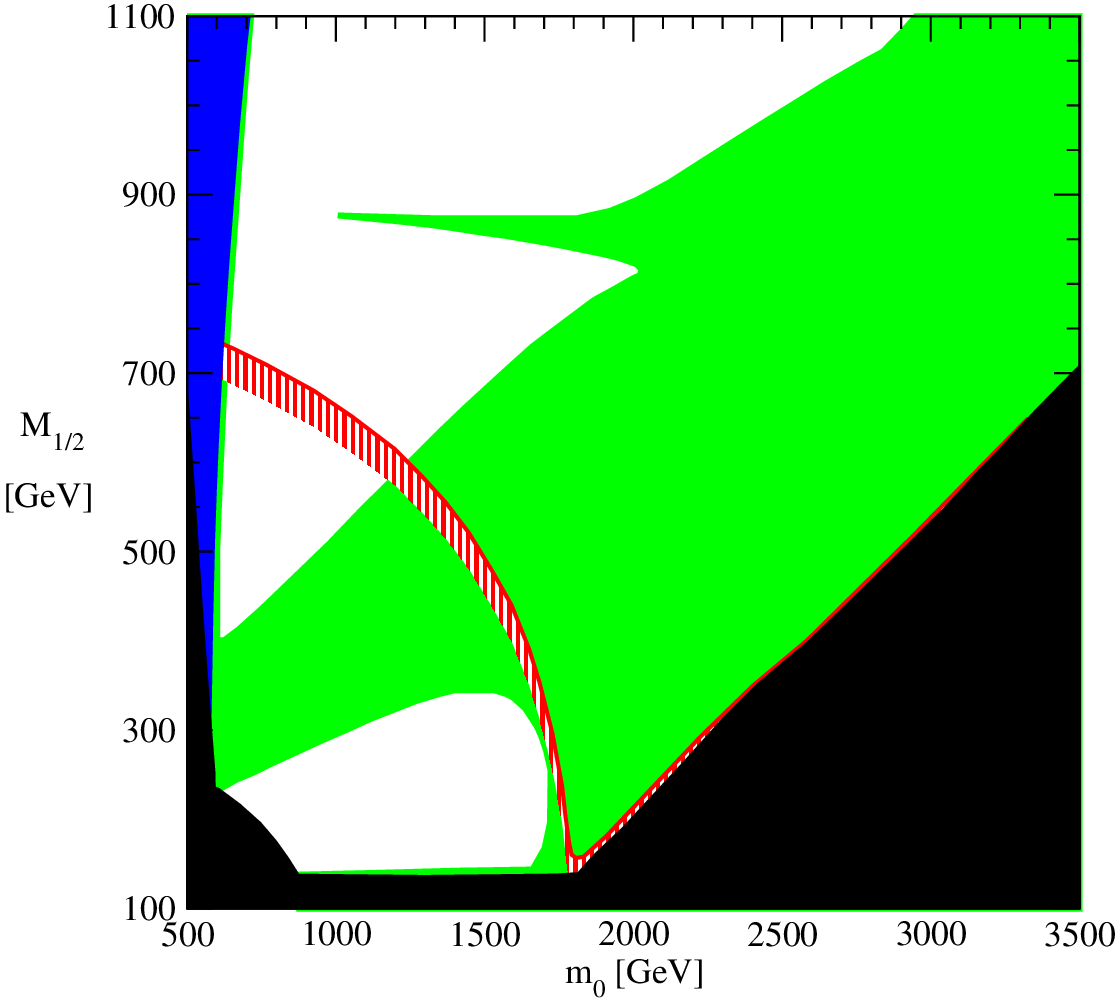, width=8cm}\hspace*{-6mm}(b)
\vspace*{-3mm}\caption{\small The WMAP allowed region in the $\m, \mhf$
plane for
(a) $\l=.1$, $\tb=50$, $\A=-1500$~GeV, $\Ak=-50$~GeV and
(b) $\l=.1$, $\tb=50$, $\A=1500$~GeV, $\Ak=250$~GeV.
Same color code as in fig.~\ref{fig:tb2}}
\label{fig:tb50}
\end{figure}

If $\A, \Ak < 0$, DM annihilation mechanisms are sfermion
coannihilation and Higgs exchange annihilation. This is
illustrated in fig.~\ref{fig:tb50}(a) for $\A = -1500$~GeV, $\Ak =
-50$~GeV. The possible Higgs resonances are again the light scalar
doublet $h$ at low $\mhf \sim 130$~GeV, just above the chargino
exclusion limit from LEP, or the pseudoscalar singlet $P$ for
slightly larger values of $\mhf \sim 200$~GeV. For $2 \lsim \m
\lsim 2.5$~TeV and $130 \lsim \mhf \lsim 200$~GeV one also finds a
WMAP allowed band where the bino LSP rapidly annihilates through
the pseudoscalar singlet resonance.
The special features of the region $\m \gsim 2.5$~TeV will be
discussed in the next subsection. The LEP constraints on the Higgs
sector excludes the region at small $\mhf$ where $.9 \lsim \m
\lsim 1.6$~TeV. For $\m \sim 750$~GeV and $130 \lsim \mhf \lsim
200$~GeV, the bino LSP coannihilates with the stop NLSP. For
larger values of $\mhf$, one also finds a bino-stau coannihilation
thin band along the forbidden stau LSP region. Smaller values of
$\m \lsim 750$~GeV are excluded as they would lead to negative
sfermion masses.

When $\A$ and $\Ak$ are both positive, all Higgs states, except the
scalar singlet $S$, can be light and contribute significantly to the bino LSP
annihilation. In fig.~\ref{fig:tb50}(b), we consider $\A = 1500$~GeV and
$\Ak = 250$~GeV. First we focus on the region $\m \lsim 2.5$~TeV.
Larger values of $\m$ will be treated in the next subsection.
The WMAP compatible regions correspond to the resonances of the light
scalar doublet $h$ just above the chargino exclusion limit from LEP at
$\mhf \sim 130$~GeV, of the heavy scalar/pseudoscalar doublets  $H/A$
in a wide band for $130 \lsim \mhf \lsim 800$~GeV, and of the pseudoscalar
singlet $P$ at still larger values of $\mhf \sim 900$~GeV.
Note however that the regions around the lighter resonances are
ruled out by the LEP limit on the Higgs sector. The annihilation channels
near the $P$ resonance are into $bb$, $\tau\tau$, as well as $WH^\pm$,
$hA$ or $ZH$. Of course coannihilation with sfermions is always possible
when $\m$ is near its lower bound.

\subsubsection{Large $\m$}\label{sec:m0}

In the CMSSM, the interest from the point of view of compatibility
with WMAP of the large $\m$ region has been widely stressed.
The main reason is that the parameter $\mu$ decreases sharply
as $\m$ increases, before entering an unphysical region where
$\mu^2 < 0$. When $\mu < M_1$, the LSP has an important
higgsino component and can annihilate efficiently into $W$ pairs.

In the CNMSSM, $\mu$ also decreases at large $\m$, although one
needs to take into account another factor: the squared mass of the
lightest pseudoscalar Higgs state can become negative at large $\m$,
thus leading to an unphysical region before the higgsino becomes LSP.
This is precisely what happens in the example we have considered
previously with $\tb=10$, in fig.~\ref{fig:tb5_tb10}(b). Large values of
$\m$ are excluded as they would lead to a negative mass squared in
the Higgs sector and the higgsino component of the LSP remains
well below 10\% over the parameter space of the theoretically
allowed region, so the annihilation into $W$ pairs is not efficient and the
relic density is too large.

For $\tb=50$ and $\Ak < 0$, as displayed in fig.~\ref{fig:tb50}(a),
the higgsino component of the LSP is also small over the allowed
parameter space at $\m \gsim 2.5$~TeV. On the other hand annihilation
through the heavy Higgs doublet $H/A$ exchange is efficient, especially
considering the $\tb$ enhanced couplings to $bb$ and $\tau\tau$.
This leads to an allowed band at large $\m$ along the boundary of
the theoretically excluded region. Note that there is a very narrow region
at this boundary, hardly distinguishable given the scale of the figure,
where the lightest pseudoscalar is excluded by LEP, just before its mass
squared becomes negative.

Finally, in the case $\tb=50$, $\Ak > 0$, fig.~\ref{fig:tb50}(b), the higgsino
component of the LSP can be large at $\m \gsim 2.5$~TeV near the
theoretically excluded region, i.e. in the lower part of the wide WMAP
allowed band. In this region, the annihilation channels are typical of a
higgsino LSP: $WW$, $bb$ as well as coannihilation channels
with the charginos or with heavier neutralinos.
In the upper part of the wide WMAP allowed band at large $\m$, annihilation
is dominated by the exchange of the heavy scalar/pseudoscalar doublet $H/A$,
as well as the pseudoscalar singlet $P$. The annihilation products include a
variety of channels such as  $bb$, $\tau\tau$, $WH^\pm$, $hA$ or $ZH$.
Note that in fig.~\ref{fig:tb50}(b) it is possible to distinguish the thin region
where the lightest pseudoscalar is excluded by LEP above the theoretically
excluded region where its mass squared becomes negative.

\subsection{Small $\l$: singlino LSP}

Next we focus on the regions of parameter space where a singlino LSP can
be found. As already mentioned, this scenario appears only for $\l \ll 1$, i.e.
$\l=.01$ in our selected scans.
Qualitatively the results are similar for smaller values of $\l$. In this case,
we have an effective MSSM with an almost decoupled  singlet sector and
we do not expect to have singlet Higgs resonances.

\subsubsection{Intermediate $\tb$}

\begin{figure}[t]
\hspace*{-2mm}\epsfig{file=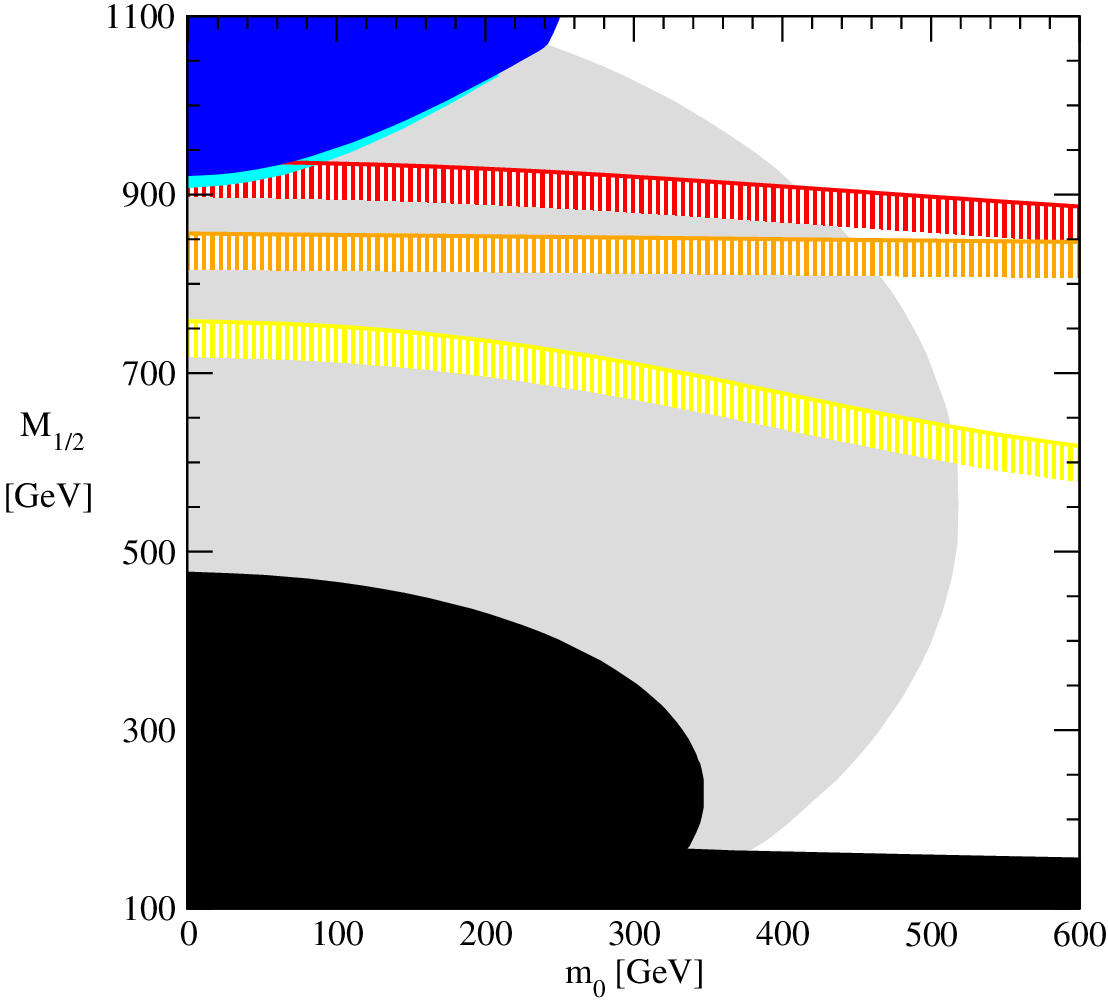, width=8cm}\hspace*{-6mm}(a)
\hspace*{6mm}\epsfig{file=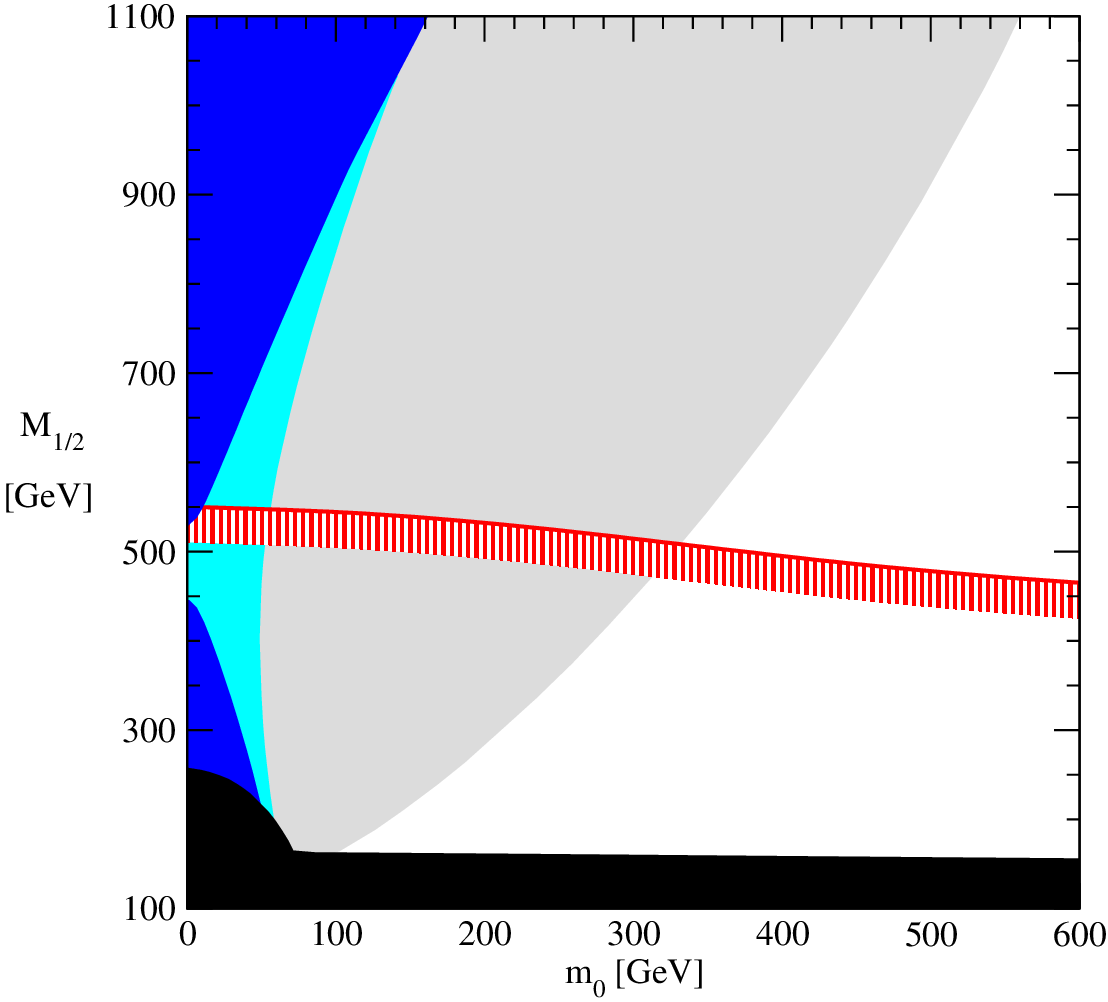, width=8cm}\hspace*{-6mm}(b)
\vspace*{-3mm}\caption{\small The WMAP allowed region in the $\m$, $\mhf$ plane (cyan) for
(a) $\l=.01$, $\tb=5$, $\A=200$~GeV, $\Ak=-10$~GeV and
(b) $\l=.01$, $\tb=10$, $\A=-20$~GeV, $\Ak=-50$~GeV.
We show the region excluded by theoretical constraints or by LEP searches on sparticles (black),
the region where the stau is the LSP (blue), the region where the singlino is the LSP (grey),
the LEP exclusion on the Higgs sector (red/hatch).
In (a) we also show the LEP constraint on the Higgs sector for $\l=.001$ (orange/hatch)
or $m_t=175$~GeV (yellow/hatch).}
\label{fig:singlino_ak-}
\end{figure}

While scanning over $\A$, $\Ak$, $\m$, $\mhf$ with $\l=.01$ and $\tb=5$ or $10$,
we have not found regions where rapid annihilation through a Higgs exchange
could take place. Thus the only mechanism that can provide the correct relic density
for a singlino LSP is coannihilation with a slepton NLSP.
This works most efficiently when the LSP and NLSP are well below the TeV scale.
We therefore expect to find WMAP allowed regions for choices of input parameters
that predict a singlino LSP at low values of $\m$ and $\mhf \lsim 1$~TeV.
This means that $\A$ cannot be too large, cf. sec.~\ref{sec:model}.

Let us start with $\Ak < 0$. As explained in sec.~\ref{sec:model},
in this case the singlino mass $m_\singlino = 2 \nu$ is positive
and grows with $\m$ and $\mhf$. Therefore, one expects to find a
singlino LSP for small values of the soft masses. For $\tb = 5$,
$\A = 200$~GeV, $\Ak = -10$~GeV, fig.~\ref{fig:singlino_ak-}(a),
the singlino is the LSP for $\m \lsim 500$~GeV, $\mhf \lsim
1100$~GeV. The singlino LSP satisfies the WMAP upper limit in
a narrow band just below the stau LSP excluded region. There,
the singlino LSP, which mass is $m_\singlino \sim 350-380$~GeV,
coannihilates with the stau and other sleptons. Most of the singlino
LSP region is excluded by the LEP constraints on the Higgs. The
whole WMAP compatible area where the LSP is a singlino can
however escape the LEP constraints for smaller values of $\l$
(e.g. $\l = .001$) or for a larger top quark mass (e.g. $m_t=175$~GeV).
From fig.~\ref{fig:singlino_ak-}(a), one can easily see the effect of $\l$
on the Higgs constraints: if one had taken $\l=.1$, the whole singlino
LSP region would have been excluded by LEP. Increasing $\tb=10$
also reduces the impact of the LEP constraint on the Higgs mass. For
example, assuming $\A = -20$~GeV, $\Ak = -50$~GeV one finds a
WMAP compatible singlino LSP region close to the the area at low
$\m$ where the stau is the LSP, see fig.~\ref{fig:singlino_ak-}(b).

\begin{figure}[t]
\hspace*{-2mm}\epsfig{file=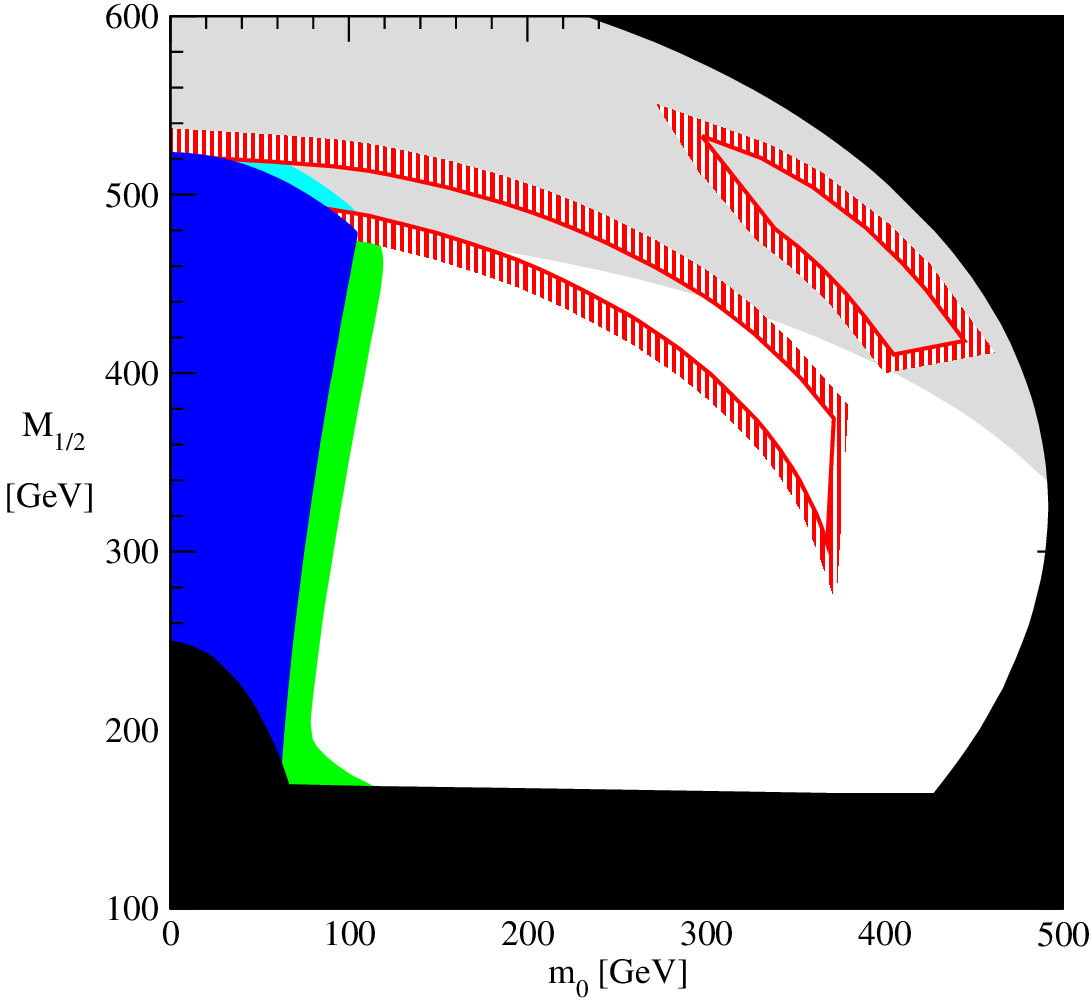,width=8cm}\hspace*{-6mm}(a)
\hspace*{6mm}\epsfig{file=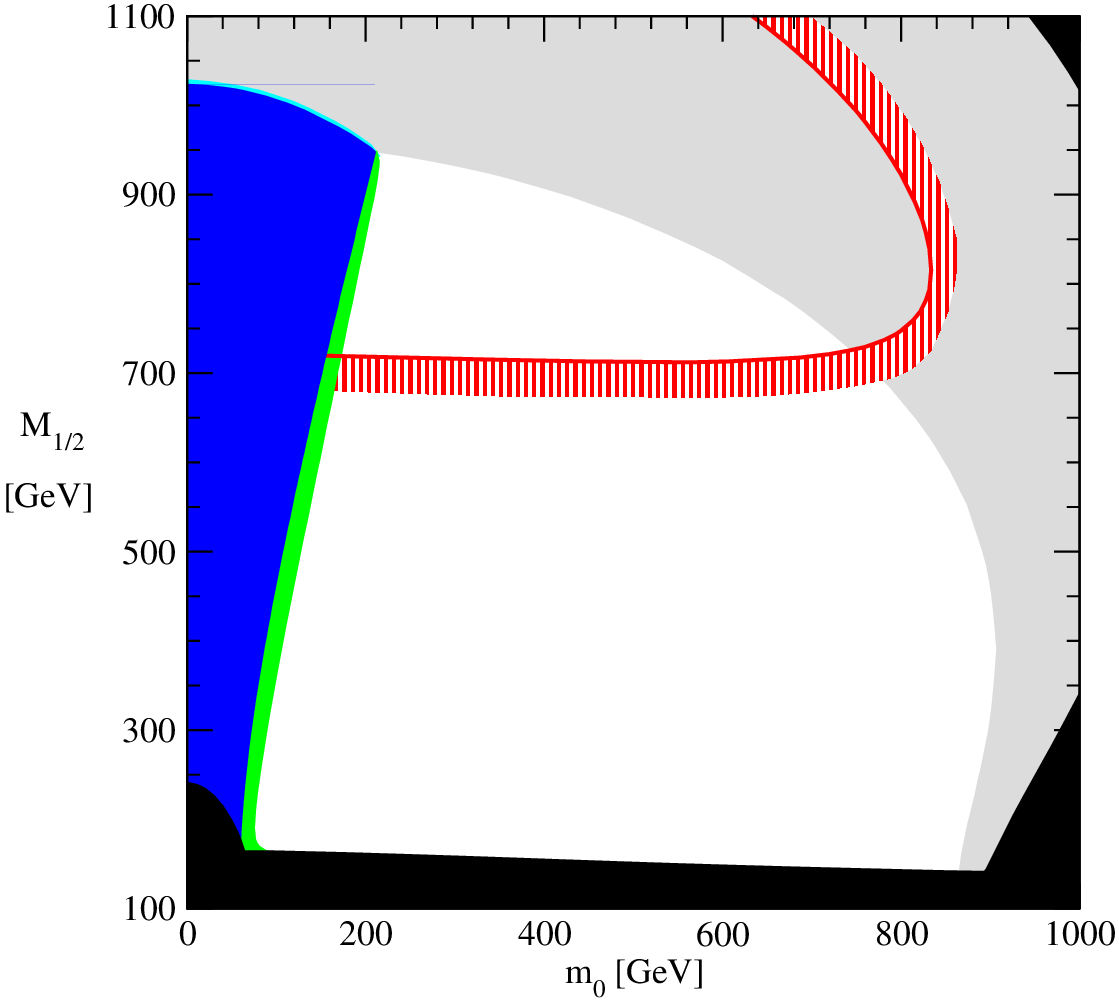,width=8cm}\hspace*{-6mm}(b)
\vspace*{-3mm}\caption{\small The WMAP allowed region in the $\m$, $\mhf$ plane
(cyan for singlino, green for non singlino LSP) for (a) $\l = .01$, $\tb = 10$,
$\A = 250$~GeV, $\Ak = 270$~GeV and
(b) $\l = .01$, $\tb = 5$, $\A = 750$~GeV, $\Ak = 10$~GeV (here $m_t=175$~GeV).
Same color code as fig.~\ref{fig:singlino_ak-}.}
\label{fig:singlino_ak+}
\end{figure}

For $\Ak > 0$, the singlino mass still grows with $\m$ and $\mhf$ but is negative.
Therefore, the singlino LSP is rather found at large values of $\m, \mhf$, just below
the excluded region where the pseudoscalar singlet mass squared becomes negative.
However, these choices of parameters always lead to light states in the Higgs sector,
excluded by LEP. Yet for $\tb = 10$, $\A = 250$~GeV, $\Ak = 270$~GeV, one
still finds a region where singlino-stau coannihilation is just enough to satisfy the WMAP
bound while LEP constraints on the Higgs are satisfied, see fig.~\ref{fig:singlino_ak+}(a).
For $\tb=5$ it is difficult to find a WMAP compatible region not excluded by LEP constraints.
However assuming $m_t=175$~GeV, one can find such an allowed region see
fig.~\ref{fig:singlino_ak+}(b). In all cases, the mass difference between the singlino and the
stau must be $\lsim 3$~GeV for the coannihilation mechanism to be efficient enough.

\subsubsection{Large $\tb$}

For $\tb = 50$, the singlino mass depends mainly on $\A$, as discussed in
sec.~\ref{sec:model}. Hence, for large values of $\mhf$ the bino is heavy and
the singlino can be the LSP. One also expects more Higgs resonances
as for $\l = .1$, $\tb = 50$ above, or in the CMSSM at large $\tb$.

\begin{figure}[t]
\hspace*{-2mm}\epsfig{file=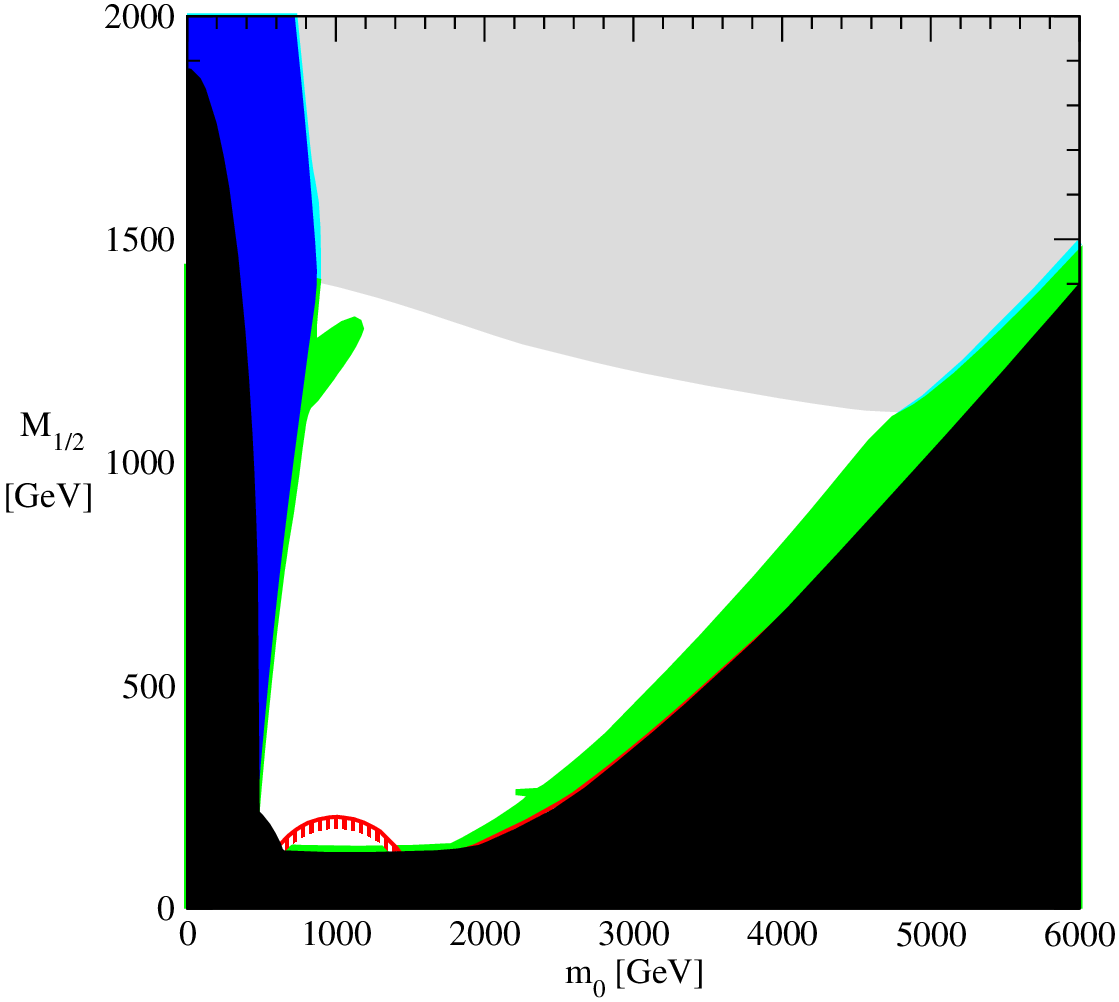, width=8cm}\hspace*{-6mm}(a)
\hspace*{6mm}\epsfig{file=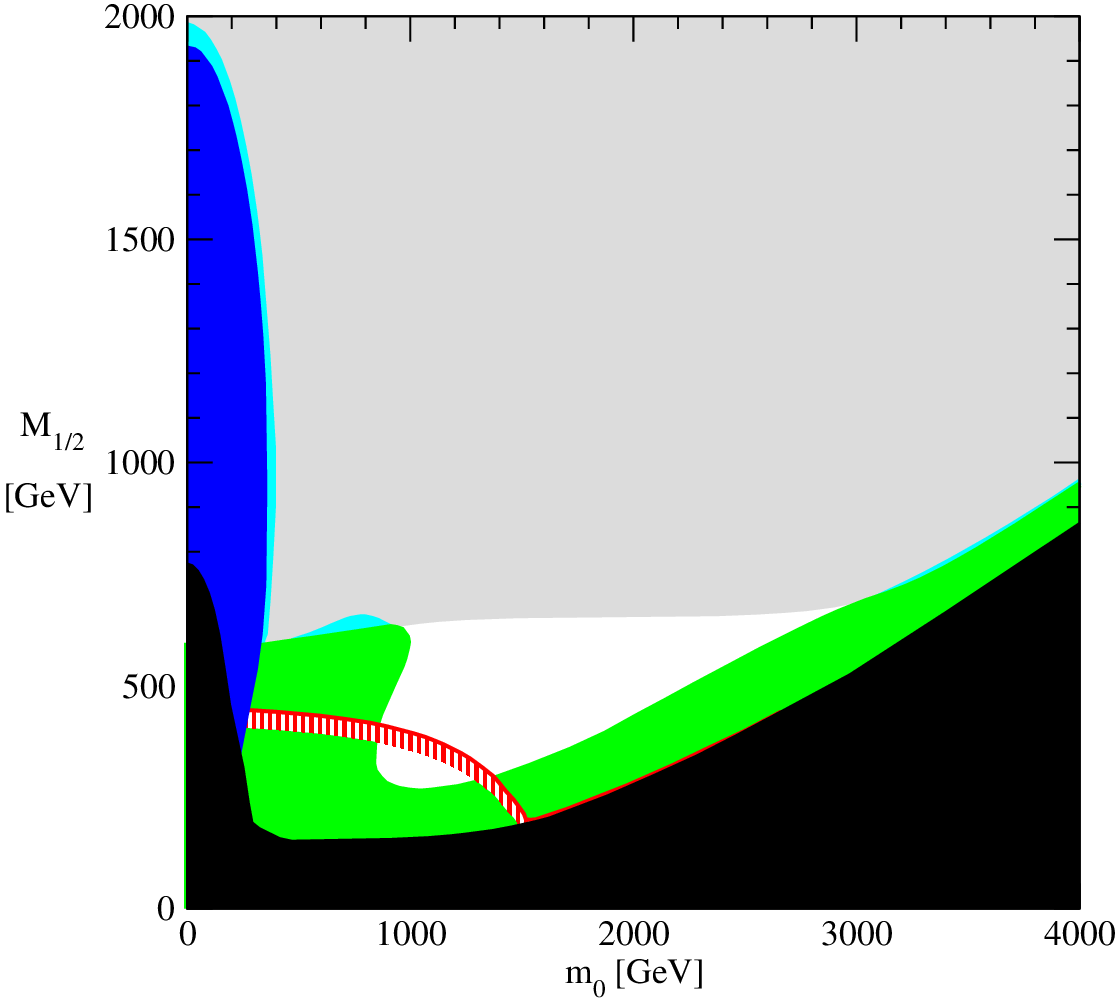, width=8cm}\hspace*{-6mm}(b)
\vspace*{-3mm}\caption{\small The WMAP allowed region in the $\m$, $\mhf$ plane for
(a) $\l = .01$, $\tb = 50$, $\A = -1000$~GeV, $\Ak = -50$~GeV and
(b) $\l = .01$, $\tb = 50$, $\A = 0$~GeV, $\Ak = 50$~GeV.
Same color code as fig.~\ref{fig:singlino_ak+}.}
\label{fig:singlino_tb50}
\end{figure}

Let us start with the case $\A = -1000$~GeV, $\Ak = -50$~GeV, as
illustrated in fig.~\ref{fig:singlino_tb50}(a). For $\mhf \lsim
1.4$~TeV, the situation is similar to the CMSSM: the LSP is mainly
bino and its relic density is below the WMAP upper bound either if
it coannihilates with the stau NLSP at low values of $\m$ (close
to the forbidden zone where the stau is the LSP) or if it
annihilates rapidly through a Higgs resonance. The light Higgs
doublet $h$ can play this role near $\mhf \sim 130$~GeV, although
this area is mostly excluded by the LEP constraints on the Higgs
sector. The heavy scalar/pseudoscalar doublet $H/A$ can also play
this role either for $\m \sim 1$~TeV, $\mhf \sim 1.3$~TeV or at
large $\m$, above the excluded region where the lightest
pseudoscalar mass squared becomes negative. When one approaches
this theoretically excluded region at very large values of $\m
\gsim 4$~TeV, the higgsino component of the LSP increases and its
relic density drops due to rapid annihilation into $W$ pairs and
coannihilation with charginos or heavier neutralinos. For $\mhf
\lsim 1.4$~TeV, the only difference with the CMSSM is the small
WMAP compatible zone at $\m \sim 2.2$~TeV, $\mhf \sim 260$~GeV
where the bino LSP rapidly annihilates though resonance of the
pseudoscalar singlet $P$. When $\mhf \gsim 1.4$~TeV, the LSP is
mainly singlino and its relic density is below the WMAP upper
bound either if it coannihilates with the stau NLSP at low values
of $\m$ or if it coannihilates with the mixed bino-higgsino NLSP
at very large values of $\m \gsim 5$~TeV. Note that the WMAP
compatible singlino LSP region at large $\m$ does not extend all
the way to the theoretically excluded region: indeed, as $\m$
increases, $\mu$ decreases and the LSP becomes predominantly
higgsino.

For $\Ak > 0$ the situation is similar. In fig.~\ref{fig:singlino_tb50}(b) we display our
results for $\A = 0$, $\Ak = 50$~GeV. Now the singlino is the LSP for $\mhf \gsim
600$~GeV and its relic density is below the WMAP upper bound in 3 different regions:
At low $\m$ where it coannihilates with the stau NLSP. For $\mhf \sim 600$~GeV and
$400 \lsim \m \lsim 1000$~GeV, where it coannihilates with the bino NLSP
which in turn annihilates rapidly through the heavy scalar/pseudoscalar doublet $H/A$
resonance. Finally, at large $\m$ the singlino LSP coannihilates with the mixed
bino-higgsino NLSP, which in turn annihilates rapidly through the $H/A$ resonance
or into typical higgsino channels, including $W$ pairs or coannihilation with charginos
and heavier neutralinos. Here again, the WMAP compatible singlino LSP region does
not extend all the way to the theoretically excluded region but stops where the LSP
becomes mainly higgsino.

\section{Discussion}\label{sec:dis}

We have achieved a first exploration of the parameter space of a constrained 
NMSSM from the point of view of DM relic density, taking into
account all theoretical and collider constraints. We have
presented our results in the $\m$, $\mhf$ plane for selected
values of $\l$, $\tb$, $\A$ and $\Ak$. We have assumed sign$(\mu)
> 0$, yet we do not expect any change in our analysis for
sign$(\mu) < 0$. We have recovered the main scenarios of the MSSM
as well as new ones.

For $\l \gsim .1$ we have shown that it was
possible to have an extra pseudoscalar singlet resonance at any
values of $\tb$. The search of this extra Higgs state might reveal
an interesting challenge at LHC~\cite{Ellwanger:2005uu}. Although
this pseudoscalar singlet state is not very heavy its couplings to
fermions are suppressed relative to a doublet pseudoscalar making
the pseudoscalar invisible unless $\tb$ is
large~\cite{Belanger:2007xxx}.

For small $\l \lsim .01$, we have shown that it was possible to have a
singlino LSP with a relic density below the WMAP upper bound.
However, such scenarios always require coannihilation with the
stau, bino or mixed bino-higgsino NLSP, and a small mass
difference between the singlino LSP and the NLSP ($\lsim 3$~GeV).
In these coannihilation scenarios, the presence of a singlino LSP
in the decay of the NLSP (sfermion or neutralino) might influence
markedly the phenomenology at colliders~\cite{Ellwanger:1997jj,Kraml:2005nx}.
The consequences for indirect detection of dark matter in all the WMAP
compatible scenarios will be analysed in a separate
publication~\cite{Hugonie:2007xxx}. These scenarios represent
only a subset of the possible scenarios for the singlino LSP DM in
the general NMSSM. Indeed, in the model with free parameters at
the weak scale, it is also possible to have annihilation of
singlino LSP through a Z or a light scalar/pseudoscalar resonance
or to have annihilation of a mixed singlino-higgsino LSP into $W$
or Higgs pairs. Such scenarios usually require $\l \gsim .1$ for which
the singlino is never the LSP in the CNMSSM\footnote{Previous phenomenological
studies of the NMSSM with strict universality at the GUT scale also concluded
that $\l \lsim .01$ was required in order to have a singlino LSP~\cite{Ellwanger:1997jj}.}.

We should also mention that if one assumes minimal flavour structure,
the branching ratios for $b\rightarrow s\gamma$ or $B_s\rightarrow \mu^+\mu^-$
might impose strong constraints on the CNMSSM parameter space, the latter being
specially relevant for large values of $\tb$~\cite{Ellwanger:2007xxx}. A global fit to
all observables including those of the flavour sector in the CNMSSM is left for a future
publication.

\section*{Acknowledgments}

We would like to thank U. Ellwanger, F. Boudjema and Y. Mambrini
for helpful discussions. This work was supported in part by
GDRI-ACPP of CNRS.


\end{document}